\documentclass[epj]{svjour}
\usepackage{graphics}
\usepackage{amsmath}
\usepackage{braket}
\usepackage[toc,page]{appendix}
\usepackage{mathtools}
\usepackage{extarrows}
\usepackage{tensor}
\usepackage[shortcuts]{extdash}

\usepackage{ucs}
\usepackage[utf8]{inputenc}
\DeclareMathOperator\arctanh{arctanh}

\newcommand{\be}{\begin{equation}}
\newcommand{\ee}{\end{equation}}
\newcommand{\bea}{\begin{eqnarray}}
\newcommand{\eea}{\end{eqnarray}}

\begin{document}
\title{Cosmogenesis as symmetry transformation}

\titlerunning{Cosmogenesis as symmetry transformation}


\author{Adam Balcerzak\inst{1,2} \and Mateusz Lisaj\inst{3}
}

\authorrunning{A. Balcerzak, M. Lisaj}
%
%
\institute{Institute of Physics, University of Szczecin,
Wielkopolska 15, 70-451 Szczecin,  Poland \and Copernicus Center for Interdisciplinary Studies, Szczepa\'nska 1/5, 31-011 Krak\'ow, Poland \and  
Institute of Mathematics, Physics and Chemistry, Maritime University of Szczecin, Wa{\l }y Chrobrego 1-2, 70-500 Szczecin, Poland}
\date{Received: date / Revised version: date}
%
\abstract{We consider the quantized bi-scalar gravity, which may serve as a locally Lorentz invariant cosmological model with varying speed of light and varying gravitational constant. The equation governing the quantum regime for the case of homogeneous and isotropic cosmological setup is a Dirac-like equation which replaces the standard Wheeler-DeWitt equation. We show that particular cosmogenesis may occur as a result of the action of the symmetry transformation which due to Wigner's theorem can either be unitary or antiunitary. We demonstrate that the transition from the pre-big-bang contraction to the post-big-bang expansion - a scenario that also occurs in string quantum cosmologies - can be attributed to the action of charge conjugation, which belongs to the class of antiunitary transformations. We also demonstrate that the emergence of the two classical expanding post-big-bang universe-antiuniverse pairs, each with opposite spin projections, can be understood as being triggered by the action of a unitary transformation resembling the Hadamard gate.
} 
\maketitle

\section{Introduction}
\label{sec:sec0}

The standpoint that quantum physics provides an accurate representation of the physical world has led to the many-worlds concept, which assumes the simultaneous existence of parallel universes \cite{Everett}. This may be the source of the quantum over classical computation advantage, in which effectively many classical calculations occur in parallel. Deutsch's concept of the multiverse \cite{Deutsch}, which extends the ideas proposed by Everett, appears to be related to the structure of the Hilbert space. The emergence of single universes occurs as a result of basis choice, which bears a resemblance to spacetime slicing in general relativity. In this slicing, a physical model distinguishes between the physical quantities defined on specific hypersurfaces and those that connect consecutive hypersurfaces. The structure of the multiverse is moreover related to the information flow, the proper description of which can be achieved by applying the theory of quantum computational networks \cite{Deutsch}. It is essential to note that the multiverse notion presented above significantly differs from the cosmological idea of the multiverse, which describes an infinite ergodic universe containing Hubble volumes and realizing all possible initial conditions, or the existence of multiple universes with varying physical constants, dimensionality, and particle content \cite{Tegmark}. However it is worth noting that the notions of quantum computation have already been utilized as an alternative language to describe the bridging of two large Anti-de-Sitter (AdS) universes \cite{Bachas}. 

Deutsch's concept of the multiverse appears to be connected with the intriguing idea that the universe itself could be considered a quantum computer \cite{Lloyd_book} although the dynamics of our universe is described by a local Hamiltonian specified by quantum field theory, which differs from the local unitary dynamics of a quantum computer. Despite this distinction, the notion that the universe behaves like a quantum computer is substantiated by the strong theoretical support of the Feynman conjecture \cite{Lloyd}, which posits that quantum computers can simulate any local system. In particular a theory of quantum gravity has been proposed, based on the principles of quantum computation, where fundamental processes are described through the framework of quantum information processing \cite{Lloyd2}.

Interesting scenarios of cosmogenesis have been found in the framework of Wheeler-DeWitt quantized low\-/energetic string cosmologies \cite{Veneziano,Gasperini}. It was also demonstrated that qualitatively similar scenarios appear in the quantized varying fundamental constants model \cite{Balcerzak1}. Considering Deutsch's idea of the multiverse, it becomes particularly tempting to view quantum cosmogenesis scenarios as quantum computational processes involving unitary transformations. Let us also note that the unitary transformations are part of a broader class of symmetry transformations, governed by Wigner's theorem \cite{Wigner}, which includes only antiunitary transformations alongside unitary ones.

In this paper, we use bi-scalar gravity theory, which may serve as a locally Lorentz invariant theory involving varying speed of light and varying gravitational constant \cite{Magueijo1}. Within the framework of this theory, various scenarios of multiverse emergence have been discovered \cite{Balcerzak2}, and the mechanisms underlying entanglement in pairs of universes have been explored \cite{Balcerzak3,Balcerzak4}. Furthermore, the utilization of the Dirac-like equation, as a replacement for the standard Wheeler-DeWitt equation in governing the quantum aspects of the model, has enabled an exploration of cosmogenesis scenarios wherein two pairs of early universe-antiuniverse emerge with differing spin projections \cite{Balcerzak5}. In the context of this particular model, which relies on the Dirac-like equation, we propose that cosmogenesis emerges as a result of the symmetry transformations. In particular, we propose two scenarios. In one, the transition from the pre-big-bang contracting phase to the post-big-bang expanding phase occurs as a result of the charge conjugation transformation, which acts as an antiunitary transformation. In the other scenario, the emergence of two classical expanding post-big-bang universe-antiuniverse pairs, each with opposite spin projections, can be attributed to the action of the unitary transformation resembling the Hadamard gate, thus depicting cosmogenesis as a basic one-gate quantum information processing.

Our paper is organized as follows. In Sec. \ref{sec:sec1}, we introduce the quantized bi-scalar gravity homogeneous and isotropic cosmological model with the Dirac-like governing equation and present its natural solutions. In Sec. \ref{sec:sec2}, we outline a scenario wherein the shift from the pre-big-bang contracting phase to the post-big-bang expanding phase happens due to the action of the charge conjugation, operating as an antiunitary transformation. In Sec. \ref{sec:sec3}, we consider the emergence of two classical expanding post-big-bang universe-antiuniverse pairs, each with opposite spin projections, as a result of a unitary transformation analogous to the Hadamard gate. In Sec. \ref{sec:conc} we give our conclusions.

\section{Quantum cosmological setup for non-minimally coupled bi-scalar gravity}  
\label{sec:sec1}
We will be proceeding in the framework of the non\-/minimally coupled bi-scalar gravity that (for the derivation of the model see App. \ref{app:1}), as it was pointed out in \cite{Magueijo1}, may serve as a theory that describes the variation of the speed of light and the gravitational constant. The consequences of quantization of such a model in the case of homogeneous and isotropic cosmological setup have already been explored in \cite{Balcerzak1,Balcerzak2,Balcerzak3,Balcerzak4,Balcerzak5} where it was shown that such models involve interesting scenarios of cosmogenesis. In particular, two different approaches have been investigated - the standard one involving Wheeler-DeWitt equation and the other one where the Wheeler-DeWitt equation was replaced by a Dirac-like equation and the wave function acquired spinorial degrees of freedom. The second approach will be adapted in what follows. 
Let us note that establishing a positively definite probability density for the Wheeler-DeWitt equation, being a hyperbolic partial differential equation, is not obvious. The resolution of this problem involved the introduction of a Dirac-square root formulation of the Wheeler-DeWitt equation \cite{Mallett,Kim,Death,Yamazaki}  resulting in ambiguities  associated with factor ordering. Another approach was to employ supersymmetric quantum mechanics \cite{Hojman,moniz1,moniz2,moniz3,moniz4,moniz5}.

The Dirac-like equation obtained for such a non\-/minimally coupled bi-scalar gravity homogeneous and isotropic cosmological model is \cite{Balcerzak5}: 
\bea
\label{dirac_expl}
\nonumber
\left[\gamma^0 \left(\partial_\eta -\frac{1}{4r}\right) +\gamma^1 \partial_1  + \gamma^2 \partial_2  + \sqrt{\bar\Lambda} e^{-\frac{\eta}{r}}\gamma^3 \partial_3 \right] \Psi =0,\\
\eea
where $\gamma^\alpha$ with $\alpha=0,1,2,3$ are the Dirac matrices, $\partial_\eta \equiv \frac{\partial}{\partial \eta}$, $\partial_i \equiv \frac{\partial}{\partial x_i}$ with $i=1,2,3$ and $\eta$, $x_i$ being variables parametrizing the extended minisuperspace dimensions, while $r$ and $\bar\Lambda$ are some parameters of the model (for the detailed description of the minisuperspace parametrization and the derivation of (\ref{dirac_expl})  see Apps. \ref{app:1} and \ref{app:2}). It is particularly interesting that such a Dirac-like equation have been obtained by utilizing the Eisenhart-Duval \cite{Eisenhart,Duval,Finn,Kan1,Kan2} method which rests on a geometrization of the potential term by extending the original minisuperspace of the model, thus making the wave function propagation geodesic. We will be looking for a solutions which fulfill the following condition:
\be
\label{reduction}
\frac{1}{i} \frac{\partial}{\partial x_3} \Psi=\Psi,
\ee
which takes us back to the physical minisuperspace (spanned by the physical degrees of freedom of the model).

The natural  solution of the Dirac-like equation (\ref{dirac_expl}) fulfilling the condition (\ref{reduction}) is \cite{Balcerzak5}:
\be
\label{spinor}
\Psi = \frac{1}{\sqrt{2}}
\begin{pmatrix}
\phi_1+\varphi_1 \\
\phi_2+\varphi_2 \\
\phi_1-\varphi_1 \\
\phi_2-\varphi_2\\
\end{pmatrix}, 
\ee
where
\begin{flalign}
\nonumber
\phi_1 =&e^{i \vec{k} \cdot \vec{x}} e^{-i \alpha} \alpha^{i k r -\frac{1}{4}}\left(C_1 U(ikr+1, 2ikr+1, 2 i \alpha) \right. &\\
&\left.  +C_2 L^{2 i k r}_{-ikr-1}(2i\alpha)\right), &\\
\nonumber
\phi_2 =&e^{i \vec{k} \cdot \vec{x}} e^{-i \alpha} \alpha^{i k r -\frac{1}{4}}\left(C_3 U(ikr, 2ikr+1, 2 i \alpha) \right. &\\
&\left. +C_4 L^{2 i k r}_{-ikr}(2i\alpha)\right), &\\
\nonumber
\varphi_1 =&e^{i \vec{k} \cdot \vec{x}} e^{-i \alpha} \alpha^{i k r -\frac{1}{4}}\left(C_5 U(ikr, 2ikr+1, 2 i \alpha) \right. &\\
&\left. +C_6 L^{2 i k r}_{-ikr}(2i\alpha)\right), &\\
\nonumber
\varphi_2=&e^{i \vec{k} \cdot \vec{x}} e^{-i \alpha} \alpha^{i k r -\frac{1}{4}}\left(C_7 U(ikr+1, 2ikr+1, 2 i \alpha) \right. &\\
&\left. + C_8 L^{2 i k r}_{-ikr-1}(2i\alpha)\right).
\end{flalign}
Here, $\vec{k}=(k_1,k_2)$, $\vec{x}=(x_1,x_2)$, $k=\sqrt{k_1^2+k_2^2}$, $U$ and $L$ denote the confluent hypergeometric function of the second kind and the associated Laguerre polynomial, respectively, $\alpha=r\sqrt{\bar\Lambda}e^{-\frac{\eta}{r}}$ while $\{C_1, C_2, C_3, C_4, C_5, C_6, C_7, C_8\}$ is a set of integration constants.

The solution of (\ref{dirac_expl}) given by (\ref{spinor}) expressed in the new representation defined by the unitary matrix:
\be
\label{fw}
F=H \otimes I=\frac{1}{\sqrt{2}}
\begin{pmatrix}
I & I  \\
I & -I \\
\end{pmatrix},
\ee
where $H=\frac{1}{\sqrt{2}}\begin{pmatrix}
1& 1  \\
1 & -1 \\
\end{pmatrix}$ is a matrix representation of the Hadamard gate and $I$ is a unit $2\times2$ matrix, reads as:
\be
\label{spinor_fw}
\Psi_F= F \Psi=
\begin{pmatrix}
\phi_1 \\
\phi_2\\
\varphi_1\\
\varphi_2\\
\end{pmatrix}.
\ee
Notice that the unitary transformation $F$ leaves untouched the four-momentum operator and the $x_3$-axis projection spin operator $\Sigma_{3}= \frac{1}{2}
\begin{pmatrix}
\sigma_3& 0  \\
0& \sigma_3\\
\end{pmatrix}$, where $\sigma_3$ stands for the Pauli matrix. The eq. (\ref{spinor_fw}) can be used to build the low-curvature ($\eta\rightarrow-\infty$ or equivalently $\alpha\equiv r\sqrt{\bar\Lambda}e^{-\frac{\eta}{r}}\rightarrow\infty$) eigenstates of both the $\Sigma_{3}$ and the four\-/momentum operator, each separately solving eq. (\ref{dirac_expl}) (see App. \ref{app:1} for the description of the low- and the high-curvature regimes). The explicit form of such low-curvature eigenstates is \cite{Balcerzak5}: 
\be
\label{pu}
\Psi_{+,+\frac{1}{2}}=A e^{i \vec{k} \cdot \vec{x}} e^{-i \alpha} \alpha^{i k r -\frac{1}{4}}\begin{pmatrix}
L^{2 i k r}_{-ikr-1}(2i\alpha) \\
0\\
U(ikr, 2ikr+1, 2 i \alpha) \\
0\\
\end{pmatrix}, 
\ee
\be
\label{pd}
\Psi_{+,-\frac{1}{2}}=B e^{i \vec{k} \cdot \vec{x}} e^{-i \alpha} \alpha^{i k r -\frac{1}{4}}\begin{pmatrix}
0 \\
U(ikr, 2ikr+1, 2 i \alpha)\\
 0\\
L^{2 i k r}_{-ikr-1}(2i\alpha)\\
\end{pmatrix},
\ee
\be
\label{nu}
\Psi_{-,+\frac{1}{2}}=C e^{i \vec{k} \cdot \vec{x}} e^{-i \alpha} \alpha^{i k r -\frac{1}{4}}\begin{pmatrix}
U(ikr+1, 2ikr+1, 2 i \alpha) \\
0\\
L^{2 i k r}_{-ikr}(2i\alpha)\\
0\\
\end{pmatrix},
\ee
\be
\label{nd}
\Psi_{-,-\frac{1}{2}}=D e^{i \vec{k} \cdot \vec{x}} e^{-i \alpha} \alpha^{i k r -\frac{1}{4}}\begin{pmatrix}
0 \\
L^{2 i k r}_{-ikr}(2i\alpha)\\
 0\\
U(ikr+1, 2ikr+1, 2 i \alpha)\\
\end{pmatrix},
\ee
where $A$, $B$, $C$ and $D$ are integration constants. The functions above satisfy in the low-curvature limit for $\eta\rightarrow-\infty$ (equivalently for $\alpha\rightarrow\infty$) the the following conditions:
\be
\label{pos_freq}
i \frac{\partial}{\partial \eta} \Psi_{+,\pm\frac{1}{2}}=\pi_{\eta(-\infty)}\Psi_{+,\pm\frac{1}{2}},
\ee
\be
\label{neg_freq}
i \frac{\partial}{\partial \eta} \Psi_{-,\pm\frac{1}{2}}=-\pi_{\eta(-\infty)}\Psi_{-,\pm\frac{1}{2}},
\ee
where $\pi_{\eta(-\infty)}= \sqrt{\bar{\Lambda}} e^{-\frac{\eta}{r}}$, which means that they are indeed the low-curvature limit eigenstates of the four\-/momentum operator. Let us also notice that (\ref{pu}), (\ref{nu}) and (\ref{pd}), (\ref{nd}) formally represent the spin $1/2$ and $-1/2$ states (along the $x_3$ axis), respectively. 

\section{Cosmogenesis as the charge conjugation transformation}
\label{sec:sec2}
Since $\pi_{\eta(-\infty)}$ and $-\pi_{\eta(-\infty)}$ correspond to the pre-big-bang collapsing branch and the expanding post-big-bang branch, respectively, of the classical solution in the low-curvature limit (see formula (\ref{low_curv}) in App. \ref{app:1}), we recognize that the solution $\Psi_{+,\pm\frac{1}{2}}$ is peaked over the pre-big-bang collapsing branch while the solution $\Psi_{-,\pm\frac{1}{2}}$ is peaked over the expanding post-big-bang branch in that limit.

Now let us define a boundary condition by assuming that initially in the high-curvature limit for $\eta\rightarrow\infty$ ($\alpha\rightarrow0$) the spinor wave function of the universe is given by $\Psi_{+,\pm\frac{1}{2}}$. In other words the in-state represents a collapsing pre-big-bang branch of the classical solution in the low-curvature limit (long before the big-bang). However, in the high-curvature limit $\eta\rightarrow\infty$ ($\alpha\rightarrow0$) the in-state $\Psi_{+,\pm\frac{1}{2}}$ is not peaked over any classical trajectory (pure quantum regime).

The charge conjugation transformation, considered a canonical example of antiunitary transformations, is defined as:
\be
\label{CC}
\overline{\Psi}\equiv\gamma^2\Psi^*.
\ee
It can be easily shown that both $\Psi$ and its charge conjugate $\overline{\Psi}$ satisfy the Dirac-like equation (\ref{dirac_expl}). We also notice that in the low-curvature limit for $\eta\rightarrow-\infty$ ($\alpha\rightarrow\infty$)
\be
\label{CCquant_genes}
\overline{\Psi}_{+,\pm\frac{1}{2}} \equiv \gamma^2 \Psi_{+,\pm\frac{1}{2}}^* \sim \Psi_{-,\mp\frac{1}{2}},
\ee
where $\sim$ indicates the physical equivalence of the wave functions in terms of the quantum numbers (related to the frequency and the spin orientation). Considering the above formula, the transition from the contracting pre-big-bang phase to the expanding post-big-bang phase ('bypassing' the big bang singularity) is made feasible through the action of the charge conjugation transformation.

\section{Cosmogenesis as the one-gate quantum information processing}
\label{sec:sec3}
The action of the unitary transformation (\ref{fw}) involving a matrix representation of the Hadamard gate on the natural solution of the Dirac-like equation (\ref{dirac_expl}) given by (\ref{spinor}) leads to the following spinor wave function (with suitable choice of the integration constants $C_i$, $i=1,2,...,8$) \cite{Balcerzak5}:
\bea
\label{spinor_general}
\nonumber 
\Psi_{F}&=&e^{i \vec{k} \cdot \vec{x}} e^{-i \alpha} \alpha^{i k r -\frac{1}{4}}\times \\ 
 \nonumber
&\times&\begin{pmatrix}
A ~ L^{2 i k r}_{-ikr-1}(2i\alpha) + C~ U(ikr+1, 2ikr+1, 2 i \alpha) \\
B ~U(ikr, 2ikr+1, 2 i \alpha)+ D ~L^{2 i k r}_{-ikr}(2i\alpha)\\
A~U(ikr, 2ikr+1, 2 i \alpha)+C~L^{2 i k r}_{-ikr}(2i\alpha) \\
B~L^{2 i k r}_{-ikr-1}(2i\alpha)+D~ U(ikr+1, 2ikr+1, 2 i \alpha)\\
\end{pmatrix}=\\
&=& \Psi_{+,+\frac{1}{2}} + \Psi_{+,-\frac{1}{2}}+\Psi_{-,+\frac{1}{2}}+\Psi_{-,-\frac{1}{2}}.
\eea
Let us note that, in the low-curvature limit as $\eta\rightarrow-\infty$ (or equivalently, as $\alpha\rightarrow\infty$),  the solutions (\ref{pu}) and (\ref{nu}) can be interpreted as post-big-bang expanding positive and negative frequency modes, respectively, with spin $1/2$ along the $x_3$ axis. Similarly, in the same low-curvature limit, the solutions (\ref{pd}) and (\ref{nd}) can also be interpreted as post-big-bang expanding positive and negative frequency modes, respectively, with spin $-1/2$ along the $x_3$ axis \cite{Balcerzak5}.

In view of formula (\ref{spinor_general}) and the interpretation given above, the emergence of the two classical expanding post-big-bang universe-antiuniverse pairs with opposite spin projections can be seen as a result of the action of the unitary matrix $F$. Since this result is an equal superposition state, the action of the unitary matrix $F$ can be regarded as analogous to the action of the Hadamard gate, thus depicting the cosmogenesis scenario as a simple one-gate quantum information processing.

\section{Conclusions}
\label{sec:conc}
We have shown that, within the framework of the bi-scalar gravity that can serve as a locally Lorentz invariant theory involving varying speed of light and varying gravitational constant with the Dirac-like equation governing its quantum regime, the cosmogenesis can be understood as a result of the action of the specific symmetry transformation. Within our analysis, we have presented two distinct scenarios. In the first scenario, the shift from the pre-big-bang contracting phase to the post-big-bang expanding phase may be treated as caused by the charge conjugation which belongs to a class of antiunitary transformations. In the alternative scenario, the appearance of two classical expanding post-big-bang universe-antiuniverse pairs (each with opposite spin projections) can be associated with the operation of a unitary transformation similar to the Hadamard gate, thereby portraying cosmogenesis as a simple quantum computation involving one gate. In this specific context, it is worth noting that the transition between successive aeons in the Conformal Cyclic Cosmology (CCC) model of the universe becomes viable through the 'intervention' of the reciprocal hypothesis \cite{Penrose} within the so called bandage region where the spacetimes defining the two consecutive aeons are solely determined by their respective lightcone structures.  Let us also note that the matter creation accompanied by the spacetime bending which gives rise to homogeneous and isotropic universes can be attributed to the action of conformal transformation \cite{Garecki} being a central notion of the mentioned reciprocal hypothesis in CCC models. However, considering the action of any antiunitary transformation as a physical process lacks strong justification. The requirement for such a transformation in the initial scenario, which leads to the transition between pre- and post-big-bang phases, might be linked to the separation of both phases by a curvature singularity, signifying a genuine spacetime singularity. It is conceivable that establishing a true physical process would require the regularization of the singularity at the classical level. In contrast, the absence of a curvature singularity in the second scenario permits the representation of the process through a unitary transformation that holds physical significance.

\appendix

\renewcommand{\theequation}{A.\arabic{equation}}
\renewcommand{\thefigure}{A\arabic{figure}}

\setcounter{equation}{0}
\setcounter{figure}{0}

\begin{appendices}

\section{Bi-scalar non-minimally coupled varying $c$ and $G$ model}
\label{app:1}
\begin{figure}
\begin{center}
\resizebox{0.4\textwidth}{!}{\includegraphics{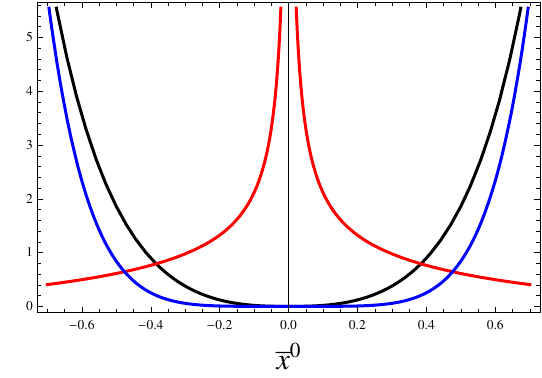}}
\caption{\label{acg} The changes in the scale factor $a$ (black), the speed of light $c$ (red), and the gravitational constant $G$ (blue) over time are demonstrated for two separate periods: $\bar{x}^0 < 0$, which pertains to the phase before the curvature singularity, and $\bar{x}^0 > 0$, representing the phase after it.}
\end{center}
\end{figure}

The bi-scalar non-minimally coupled gravity theory which can be use to model variation in the speed of light $c$ and the gravitational constant $G$ \cite{Magueijo1} is given by the following action \cite{Balcerzak1}:
\begin{equation}
\label{action}
S=\int \sqrt{-g}  \left(\frac{e^{\phi}}{e^{\psi}}\right) \left[R+\Lambda + \omega (\partial_\mu \phi \partial^\mu \phi + \partial_\mu \psi \partial^\mu \psi)\right] d^4x,
\end{equation}
where $\phi$ and $\psi$ stand for non-minimally coupled scalar fields, $R$ corresponds to the Ricci scalar, $\Lambda$ denotes the cosmological constant, and $\omega$ is as a parameter of the model. This action can be derived by substituting in the original Einstein-Hilbert action the following functions for $c$ and $G$:
\bea
c^3&=&e^{\phi}, \\
G&=&e^\psi.
\eea
The following fields transformation:
\begin{eqnarray}
\label{fred}
\phi &=& \frac{\beta}{\sqrt{2\omega}}+\frac{1}{2} \ln \delta, \\
\psi &=& \frac{\beta}{\sqrt{2\omega}}-\frac{1}{2} \ln \delta
\end{eqnarray}
brings the action (\ref{action}) to the Brans-Dicke form:
\bea
\label{actionBD} \nonumber
S=\int \sqrt{-g}\left[ \delta (R+\Lambda) +\frac{\omega}{2}\frac{\partial_\mu \delta \partial^\mu \delta}{\delta} + \delta \partial_\mu \beta \partial^\mu \beta\right]d^4x.\\
\eea
The variability of the speed of light with respect to space and time variables results in the general covariance breaking. To overcome this problem, a particular reference frame known as the 'light frame' must be chosen. This frame serves as the primary reference for constructing the model. Following the approach detailed in \cite{Albrecht,Barrow1}, we designate the cosmological frame as the light frame for our model, defined by the FLRW metric:
\be
\label{FLRW}
ds^2=-N^2(dx^0)^2+ a^2(dr^2 + r^2 d\Omega^2),
\ee
where $N$ denotes the lapse function and $a$ is the scale factor (both functions depend on the coordinate $x^0$). 
The action (\ref{actionBD}) expressed in the light frame (\ref{FLRW}) is:
\begin{eqnarray}
\label{action_sym} \nonumber
S &=& \frac{3 V_0}{8 \pi} \int dx^0 \left(-\frac{a^2}{N} a' \delta' - \frac{\delta}{N} a a'^2   + \Lambda \delta a^3 N  \right. \\
&-& \left.\frac{\omega}{2} \frac{a^3}{N} \frac{\delta'^2}{\delta}-\frac{a^3}{N}\delta \beta'^2 \right),
\end{eqnarray}
where $()'\equiv \frac{\partial}{\partial x^0}$. The behavior of the model  defined by (\ref{action_sym}) in the gauge  $N = a^3 \delta$ is described by its solutions given by \cite{Balcerzak1}:
\bea
\label{rozwio1}
a&=& \frac{1}{D^2 {(e^{ F x^0})}^2 \sinh ^ M |\sqrt{(A^2-9)\Lambda }x^0| },\\
\label{rozwio2}
\delta &=& \frac{D^6 {(e^{ F x^0})}^6}{\sinh ^ W |\sqrt{(A^2-9)\Lambda }x^0|},
\eea
where   $A=\frac{1}{\sqrt{1-2\omega}}$, $M=\frac{3-A^2}{9-A^2}$, $W=\frac{2A^2}{9-A^2}$ while $D$ and $F$ are the integration constants. The variable $x^0$ expressed by the rescaled cosmic time $\bar{x}^0$ is \cite{Balcerzak1}: 
\begin{equation}
\label{conect}
\begin{split}
x^0 &= \frac{2}{\sqrt{(A^2-9)\Lambda}}  \arctanh \left(e^{\sqrt{(A^2-9)\Lambda}\bar{x}^0}\right)\,,
\hspace{0.2cm}
\text{for $\bar{x}^0<0$}\,,
\\
x^0 &= \frac{2}{\sqrt{(A^2-9)\Lambda}}  \arctanh \left(e^{- \sqrt{(A^2-9)\Lambda}\bar{x}^0}\right)\,,
\hspace{0.2cm}
\text{for $\bar{x}^0>0$}\,,
\end{split}
\end{equation}
where as in \cite{Balcerzak1} we limit the scope of examined models to cases with $A^2>9$. The sequence of events involves a pre-big-bang contraction, taking place for $\bar{x}^0<0$, followed by a post-big-bang expansion, occurring for $\bar{x}^0>0$. Both of these phases adhere to the solutions (\ref{rozwio1}) and (\ref{rozwio2}), supplemented by (\ref{conect}). These distinct phases are separated by a curvature singularity situated at $\bar{x}^0=0$. The changes in the fundamental constants $c$ and $G$ are contained in (\ref{rozwio1}) and (\ref{rozwio2}). In accordance with these formulas, the gravitational constant $G$ approaches zero, while the speed of light $c$ goes to infinity as the universe converges towards the curvature singularity at $\bar{x}^0=0$ (see Fig. (\ref{acg})).

By employing the field transformations given by:
\bea
\nonumber
X = \ln(a \sqrt{\delta}),&& \hspace {0.3cm} Y = \frac{1}{2A} \ln \delta,\\ \nonumber
\eta=r(AY-3X), \hspace {0.1cm} x_1&=&r(3Y-AX), \hspace {0.1cm} x_2=2\sqrt{\tilde{V}_0}\beta,\\
\eea
where $\tilde{V}_0 = \frac{3 V_0}{8\pi}$ and $r=2\sqrt{\frac{\tilde{V}_0}{A^2-9}}$, the action (\ref{action_sym}) can be reformulated in the following manner:
\be
\label{action_simple}
S= \int dx^0 \left[\frac{1}{4}(\eta'^2-x_1'^2-x_2'^2)+\bar{\Lambda}e^{-2\frac{\eta}{r}}\right],
\ee
with $\bar{\Lambda} = \tilde{V}_0\Lambda$. The Hamiltonian corresponding to the action (\ref{action_simple}) is:
\be
\label{ham}
H=\pi_\eta^2 -\pi_{x_1}^2- \pi_{x_2}^2-\bar{\Lambda}e^{-2\frac{\eta}{r}},
\ee
where $\pi_\eta=\frac{\eta'}{2}$, $\pi_{x_1}=-\frac{x_1'}{2}$ and $\pi_{x_2}=-\frac{x_2'}{2}$ denote the conjugated momenta. The structure of the Hamiltonian (\ref{ham}) indicates the constancy of both $\pi_{x_1}$ and $\pi_{x_2}$ throughout the evolution. As a result, the classical evolution can be visualized as a particle encountering an exponential potential barrier. The solutions to Hamilton's equations related to the Hamiltonian (\ref{ham}) are:
\bea
\label{ham_sol1}
\eta&=&\ln \sinh|\sqrt{(A^2-9)\Lambda }x^0|, \\
\label{ham_sol2}
x_1&=& -2 \pi_{x_1} x^0 + E, \\
\label{ham_sol3}
x_2&=&-2 \pi_{x_2} x^0 + P,
\eea
where $E$ and $P$ represent integration constants. Upon analyzing the solution (\ref{ham_sol1}), it becomes apparent that the parameter $\eta$ can be used to define two distinct situations. The first pertains to the high-curvature regime, characterized by the scale factor $a$ approaching zero as $\eta$ tends towards infinity, $\eta\rightarrow\infty$. The second situation corresponds to the low-curvature regime, characterized by higher values of the scale factor $a$ as $\eta$ approaches negative infinity, $\eta\rightarrow-\infty$. Furthermore, it can be checked that the high-curvature regime, $\eta\rightarrow\infty$, exhibits the following asymptotic values of momentum $\pi_\eta$:
 \begin{equation}
\pi_\eta = \begin{cases}
    \sqrt{\bar{\Lambda}} & \text{(collapsing pre-big-bang solution)} \\
    -\sqrt{\bar{\Lambda}} & \text{(expanding post-big-bang solution).}
\end{cases}
\end{equation}
On the other hand, in the low-curvature regime,  $\eta\rightarrow-\infty$, the asymptotic values of $\pi_\eta$ are:
\begin{equation} \label{low_curv}
\pi_\eta = \begin{cases}
    \sqrt{\bar{\Lambda}} e^{-\frac{\eta}{r}} & \text{(collapsing pre-big-bang solution)} \\
    -\sqrt{\bar{\Lambda}}e^{-\frac{\eta}{r}} & \text{(expanding post-big-bang solution).}
\end{cases}
\end{equation} 
  
The Wheeler-DeWitt equation, describing the quantum regime of the considered model, can be derived by applying the Jordan quantization rules to the Hamiltonian constraint $H\Phi=0$. In these rules, the canonical momenta are replaced with corresponding operators according to the following scheme: $\pi_\eta\rightarrow \hat{\pi}_\eta=-i \frac{\partial}{\partial \eta}$, $\pi_{x_1}\rightarrow \hat{\pi}_{x_1}=-i \frac{\partial}{\partial x_1}$ and $\pi_{x_2}\rightarrow \hat{\pi}_{x_2}= -i \frac{\partial}{\partial x_2}$. The resulting Wheeler-DeWitt equation is:
 \begin{equation}
\label{KG}
\ddot{\Phi} - \Delta \Phi + m_{eff}^2(\eta) \Phi=0,
\end{equation}
where $\dot{( )}\equiv\frac{\partial}{\partial \eta}$, $\Delta = \frac{\partial^2}{\partial x_1^2}+\frac{\partial^2}{\partial x_2^2}$ and $m_{eff}^2(\eta)= \bar{\Lambda} e^{-\frac{2}{r}\eta}$.

\section{Eisenhart-Duval lift of the bi-scalar non-minimally coupled varying $c$ and $G$ cosmological model - the Dirac-like equation}
\label{app:2}

The Eisenhart-Duval lifting scheme provides a geometrical description of the evolution of the system, even in the presence of the potential term  \cite{Eisenhart,Duval}. This is achieved by incrementing the dimensionality of the original minisuperspace of the model by one, through the addition of an auxiliary degree of freedom. In this extended minisuperspace, the trajectories of the system follow geodesics associated with the so-called lifted metric equipping the extended minisuperspace. Meanwhile, their projections onto the original minisuperspace correspond to the original trajectories of the system. We will use the Eisenhart-Duval lift adjusted the cosmological model arising form the scalar-tensor gravity  \cite{Finn,Kan1,Kan2}. Let us consider the Lagrangian defined by the action  (\ref{action_simple}) of our bi-scalar non-minimally coupled cosmological model:
\be
\label{lag}
L=\frac{1}{4}(\eta'^2-x_1'^2-x_2'^2)+\bar{\Lambda}e^{-2\frac{\eta}{r}}.
\ee
The so-called lifted Lagrangian associated with (\ref{lag}) that gives rise to the extended minisuperspace is:
\be
\label{lag_ext}
L_{ext}= \frac{1}{2}\left( \frac{\eta'^2}{2} - \frac{x_1'^2}{2} - \frac{x_2'^2}{2} - \frac{x_3'^2}{2 \bar{\Lambda}e^{-2\frac{\eta}{r}}}\right),
\ee
where $x_3$ is an axillary dynamical variable which extends the original minisuperspace of the model. The lifted metric on the extended minisuperspace is \cite{Balcerzak5}:
\be
\label{metric_ext}
\tilde{G}_{\alpha\beta}=diag\left(\frac{1}{2},-\frac{1}{2},-\frac{1}{2}, \frac{-1}{2 \bar{\Lambda}e^{-2\frac{\eta}{r}}}\right).
\ee
Due to the conformal invariance of the Hamiltonian constraint given by:
\be
\label{ham_const}
\frac{1}{2} \tilde{G}^{\alpha\beta} \tilde{P}_\alpha \tilde{P}_\beta=0,
\ee
with $\tilde{P}_\alpha=\tilde{G}_{\alpha\beta} x'^\beta$, in the following, we will use the conformally equivalent extended minisuperspace metric:
\be
\label{metric_ext_conf}
G_{\alpha\beta}=\Omega^2 \tilde{G}_{\alpha\beta},
\ee
where $\Omega^2=\left[{2 \bar{\Lambda}e^{-2\frac{\eta}{r}}}\right]^\frac{1}{n-2}$ with $n$ denoting the dimension of the extended minisuperspace. Since in the case of our model $n=4$ the conformal factor is:
\be
\label{conf_fact}
\Omega^2=\sqrt{2 \bar{\Lambda}}e^{-\frac{\eta}{r}}
\ee
and the conformally equivalent extended minisuperspace metric reads:
\be
\label{the_metric}
G_{\alpha\beta}=diag\left(\sqrt{\frac{\bar{\Lambda}}{2}}e^{-\frac{\eta}{r}},-\sqrt{\frac{\bar{\Lambda}}{2}}e^{-\frac{\eta}{r}},-\sqrt{\frac{\bar{\Lambda}}{2}}e^{-\frac{\eta}{r}},\frac{-e^{\frac{\eta}{r}}}{\sqrt{2\bar{\Lambda}}}\right).
\ee
The expression for the covariant Wheeler-DeWitt equation in the extended case is:
\be
\label{WDW_cov}
\frac{1}{\sqrt{-G}}\partial_\alpha (\sqrt{-G} G^{\alpha\beta} \partial_\beta \Phi)=0,
\ee 
where $G$ denotes the determinant of $G_{\alpha\beta}$. Using  (\ref{the_metric}), the Wheeler-DeWitt equation (\ref{WDW_cov}) takes the following form:
\be
\label{WDW_ext}
\partial_\eta^2 \Phi - \partial_1^2 \Phi - \partial_2^2 \Phi - \bar{\Lambda} e^{-2\frac{\eta}{r}} \partial_3^2 \Phi=0.
 \ee
 The condition that leads to the reduction of the extended case (\ref{WDW_ext}) to the initial one (\ref{KG}) reads:
 \be
\label{reduce}
\partial_{3}^2 \Phi=-\Phi.
\ee

To obtain the Dirac-like cosmological equation, we enforce covariance on the minisuperspace \cite{Kan1,Kan2} as we did in the case of the Wheeler-DeWitt equation (\ref{WDW_cov}). Due to the conformal covariance of the Dirac equation without the mass term \cite{Hijazi}, we once again employ (\ref{the_metric}) as the lifted metric. The resulting Dirac-like cosmological equation is:
\be
\label{dirac} 
\hat{\gamma}^\alpha D_\alpha \Psi \equiv \gamma^A \tensor{e}{_A ^\alpha} D_\alpha \Psi=0,
\ee
where $\gamma^A$ denotes Dirac matrices:
\begin{eqnarray}
\gamma^{0} =
\begin{pmatrix}
I & 0  \\
0 & -I \\
\end{pmatrix}, \hspace{0.5 cm}
\gamma^{k} =
\begin{pmatrix}
0 & \sigma_k \\
-\sigma_k & 0 \\
\end{pmatrix}.
\end{eqnarray}
Here, $k$ takes values of 1, 2 and 3, $\sigma_k$ represents the Pauli matrices, and $I$ denotes the identity matrix. The symbols  $\tensor{e}{_A ^\alpha}$ in (\ref{dirac}) are vector fields which fulfil $\eta_{AB}=\tensor{e}{_A ^\alpha} \tensor{e}{_B ^\beta}  G_{\alpha\beta}$ while the covariant derivative $D_\alpha$ is:
\be
\label{cov_der}
D_\alpha=\partial_\alpha + \Gamma_\alpha,
\ee
where 
\be
\label{big_gamma}
\Gamma_\alpha=\frac{1}{2} \omega_{AB_\alpha} \Sigma^{AB}.
\ee
Here, $\omega_{AB_\alpha} = G_{\nu\mu} \tensor{e}{_A ^\mu} \nabla_\alpha \tensor{e}{_B ^\nu}$ denotes the spin connection and $\Sigma^{AB}=\frac{1}{4}\left[\gamma^A,\gamma^B \right]$. The expression for the particular vector fields $\tensor{e}{_A ^\alpha}$ takes the following form:
\be
\label{vector_fields}
\tensor{e}{_A ^\alpha} =diag \left(\left(\frac{\bar\Lambda}{2} \right)^{-\frac{1}{4}}e^{\frac{\eta}{2r}}\left(1,1,1\right), \left(2 \bar\Lambda\right)^{\frac{1}{4}}e^{-\frac{\eta}{2r}}\right),
\ee
while the non-vanishing elements of the spin connection $\omega_{AB_\alpha}$ are given by:
\begin{eqnarray}
\label{spin_conn}
\omega_{10_1}&=&\frac{1}{2r}, \hspace{0.5cm} \omega_{01_1}=-\frac{1}{2r}, \nonumber \\
\omega_{20_2}&=&\frac{1}{2r}, \hspace{0.5cm} \omega_{02_2}=-\frac{1}{2r},  \\
\omega_{30_3}&=&\frac{-e^{\frac{\eta}{r}}}{2r\sqrt{\bar\Lambda}}, \hspace{0.5cm} \omega_{03_3}=\frac{e^{\frac{\eta}{r}}}{2r\sqrt{\bar\Lambda}}. \nonumber
\end{eqnarray}
Thus, the Dirac-like cosmological equation on the extended minisuperspace is \cite{Balcerzak5}:
\bea
\label{dirac_expl_app}
\nonumber
\left[\gamma^0 \left(\partial_\eta -\frac{1}{4r}\right) +\gamma^1 \partial_1  + \gamma^2 \partial_2  + \sqrt{\bar\Lambda} e^{-\frac{\eta}{r}}\gamma^3 \partial_3 \right] \Psi =0.\\
\eea

\end{appendices}


\begin{thebibliography}{}

\bibitem{Everett}
H. Everett, "Relative State" Formulation of Quantum Mechanics,  Rev. Mod. Phys. \textbf{29}, 454  (1957)

\bibitem{Deutsch}
D. Deutsch, The Structure of the Multiverse, arXiv:quant-ph/0104033

\bibitem{Tegmark}
M. Tegmark, Parallel universes,  Sci. Am. \textbf{288}, 40-51 (2003)

\bibitem{Bachas}
C. Bachas, I. Lavdas, Quantum Gates to Other Universes, Prog. Phys. \textbf{66}, 1700096 (2018)

\bibitem{Lloyd_book}
S. Lloyd, The Universe as Quantum Computer. A Computable Universe, World Scientific, Singapore, (2012), pp. 567-581

\bibitem{Lloyd}
S. Lloyd, Universal Quantum Simulators, Science. \textbf{273}, 1073 (1996)

\bibitem{Lloyd2}
S. Lloyd, A theory of quantum gravity based on quantum computation, arXiv:quant-ph/0501135

\bibitem{Veneziano}
M. Gasperini, G. Veneziano, Birth of the Universe as quantum scattering in string cosmology, Gen. Rel. Grav. \textbf{28}, 1301 (1996) 

\bibitem{Gasperini}
M. Gasperini, Birth of the universe as antitunneling from the string perturbative vacuum, Int. J. Mod. Phys. D \textbf{10}, 15 (2001) 

\bibitem{Balcerzak1}
A. Balcerzak,  Non-minimally coupled varying constants quantum cosmologies,  JCAP \textbf{04}, 019  (2015)

\bibitem{Wigner}
E. P. Wigner,  Group Theory and its Application to the Quantum Mechanics of Atomic Spectra, Academic Press, New York, (1959), pp. 233–236 

\bibitem{Magueijo1}
J. Magueijo, Covariant and locally Lorentz-invariant varying speed of light theories,  Phys. Rev. D \textbf{62}, 103521 (2000)

\bibitem{Balcerzak2}
A. Balcerzak, K. Marosek, Emergence of multiverse in third quantized varying constants cosmologies, Eur. Phys. J. C \textbf{79}, 563 (2019) 

\bibitem{Balcerzak3}
A. Balcerzak, K. Marosek, Doubleverse entanglement in third quantized non-minimally coupled varying constants cosmologies,  Eur. Phys. J. C \textbf{80}, 709 (2020) 

\bibitem{Balcerzak4}
A. Balcerzak, M. Lisaj, Decaying universes and the emergence of Bell-type interuniversal entanglement in varying fundamental constants cosmological model,  Eur. Phys. J. C \textbf{82}, 732 (2022) 

\bibitem{Balcerzak5}
A. Balcerzak, M. Lisaj, Spinor wave function of the Universe in non-minimally coupled varying constants cosmologies,   Eur. Phys. J. C \textbf{83}, 401 (2023)  

\bibitem{Mallett} 
R. L. Mallett, Dirac quantization of Friedmann cosmologies,  Class. Quantum Grav. \textbf{12}, L1 (1995)

\bibitem{Kim}
C. M. Kim, S. K. Oh, Dirac-Square-Root formulation of some types of minisuperspace quantum cosmology, Journal of the Korean Physical Society \textbf{29}, 549 - 553 (1996)

\bibitem{Death}
P. D. D'Eath, S. W. Hawking, O. Obreg{\'o}n, Supersymmetric Bianchi models and the square root of the Wheeler-DeWitt equation, Phys. Lett. B \textbf{300}, 44-48 (1993)

\bibitem{Yamazaki}
H. Yamazaki, T. Hara, Dirac Decomposition of Wheeler-DeWitt Equation in the Bianchi Class A Models, Progress of Theoretical Physics \textbf{106}, 323–337 (2001)

\bibitem{Hojman}
S. A. Hojman, F. A. Asenjo, Supersymmetric Majorana quantum cosmologies, Phys. Rev. D \textbf{92}, 083518 (2015)

\bibitem{moniz1}
P. V. Moniz, Origin of structure in supersymmetric quantum cosmology, Phys. Rev. D \textbf{57}, R7071 – R7074 (1998)

\bibitem{moniz2}
P. V. Moniz, Supersymmetric quantum cosmology shaken, not stirred, Int. J. Mod. Phys. A \textbf{11}, 4321 – 4382 (1996)

\bibitem{moniz3}
P. V. Moniz, Quantum Cosmology - The Supersymmetric Perspective - Vol. 1: Fundamentals: Preface. Lecture Notes in Physics, vol. 803 Springer, Berlin, Heidelberg, (2010), pp. vii–viii

\bibitem{moniz4}
P. V. Moniz, Quantum cosmology - The supersymmetric perspective - Vol. 2: Advanced topics. Lecture Notes in Physics, vol. 804 Springer, Berlin, Heidelberg, (2010), pp. 1–297

\bibitem{moniz5}
C. Kiefer, L. Tobias, P. V. Moniz, Semiclassical approximation to supersymmetric quantum gravity, Phys. Rev. D \textbf{72}, 1 - 19 (2005)

\bibitem{Eisenhart}
L. P. Eisenhart, Dynamical trajectories and geodesics, Ann. Math. \textbf{30}, 591 (1928)

\bibitem{Duval}
C. Duval, G. Burdet, H. P. K{\"u}nzle, M. Perrin, Bargmann structures and Newton-Cartan
theory, Phys. Rev. D \textbf{31}, 1841 (1985)

\bibitem{Finn}
K. Finn, S. Karamitsos, Finite measure for the initial conditions of inflation, Phys. Rev. D \textbf{99}, 063515 (2019)

\bibitem{Kan1}
N. Kan, T. Aoyama, T. Hasegawa, K. Shiraishi, Eisenhart-Duval lift for minisuperspace quantum cosmology,  Phys. Rev. D \textbf{104}, 086001 (2021)

\bibitem{Kan2}
N. Kan, T. Aoyama, T. Hasegawa, K. Shiraishi, Third quantization for scalar and spinor wave functions of the Universe in an extended minisuperspace, Class. Quantum Grav. \textbf{39}, 165010 (2022)

\bibitem{Penrose}
R. Penrose, Cycles of Time: An Extraordinary New View of the Universe, Bodley Head, London (2010)

\bibitem{Garecki}
M. P. D\c{a}browski, J. Garecki, D. B. Blaschke, Conformal transformations and conformal invariance in gravitation, Annalen Phys. (Berlin) \textbf{18}, 13-32 (2009)

\bibitem{Albrecht}
A. Albrecht, J. Magueijo, Time varying speed of light as a solution to cosmological puzzles, Phys. Rev. D \textbf{59}, 043516 (1999)

\bibitem{Barrow1}
J. D. Barrow,  Cosmologies with varying light speed, Phys. Rev. D \textbf{59}, 043515 (1999)

\bibitem{Hijazi}
O. Hijazi, A conformal lower bound for the smallest eigenvalue of the Dirac operator and killing spinors, Commun. Math. Phys. \textbf{104}, 151–162 (1986)

\end{thebibliography}
\end{document}